\documentclass[twocolumn,amsmath,amssymb,aps,preprintnumbers,prd]{revtex4-1}
\usepackage{times}
\usepackage{amsmath}
\usepackage{amsfonts}
\usepackage{graphicx} % Include figure files
\usepackage{dcolumn} % Align table columns on decimal point
\usepackage{bm}         % bold math
\usepackage{color}
\usepackage{epsfig}
\usepackage{slashed}

\def\beq{\begin{equation}}
\def\eeq{\end{equation}}
\def\beqn{\begin{eqnarray}} \def\eeqn{\end{eqnarray}}

\def\beq{\begin{equation}}
\def\eeq{\end{equation}}
\def\bea{\begin{eqnarray}}
\def\eea{\end{eqnarray}}
\def\beqn{\begin{eqnarray}} \def\eeqn{\end{eqnarray}}
\def\beeq{\begin{eqnarray}}
\def\eeeq{\end{eqnarray}}

\begin{document}

%%%%%%%%%%%%%%%%%%%%%%%%%%%%%%%%%%%%%%%%%%%%%%%%%%%%%%%%%%%%%%%%%%%%%
\preprint{DESY 21-017}

\title{Geometrical approach to causality in multi-loop amplitudes}

\author{German F. R. Sborlini~$^{a}$} \email{german.sborlini@desy.de}

\affiliation{ ${}^{a}$ Deutsches Elektronen-Synchrotron DESY, Platanenallee 6, D–15738 Zeuthen, Germany.}

%\date{\today}

%%%%%%%%%%%%%%%%%%%%%%%%%%%%%%%%%%%%%%%%%%%%%%%%%%%%%%%%%%%%%%%%%%%%%%%%%%%%%%%%%%%%%%%%%%%%%%%%%%%
%%%%%%%%%%%%%%%%%%%%%%%%%%%%%%%%%%%%%%%%%%%%%%%%%%%%%%%%%%%%%%%%%%%%%%%%%%%%%%%%%%%%%%%%%%%%%%%%%%%
\begin{abstract}
An impressive effort is being placed in order to develop new strategies that allow an efficient computation of multi-loop multi-leg Feynman integrals and scattering amplitudes, with a particular emphasis on removing spurious singularities and numerical instabilities. In this article, we describe an innovative geometric approach based on graph theory to unveil the causal structure of any multi-loop multi-leg amplitude in Quantum Field Theory. Our purely geometric construction reproduces faithfully the manifestly causal integrand-level behaviour of the Loop-Tree Duality representation. We found that the causal structure is fully determined by the \emph{vertex matrix}, through a suitable definition of connected partitions of the underlying diagrams. Causal representations for a given topological family are obtained by summing over subsets of all the possible causal entangled thresholds that originate connected and oriented partitions of the underlying topology. These results are compatible with Cutkosky rules. Moreover, we found that diagrams with the same number of vertices and multi-edges exhibit similar causal structures, regardless of the number of loops.
\end{abstract}

\maketitle

%%%%%%%%%%%%%%%%%%%%%%%%%%%%%%%%%%%%%%%%%%%%%%%%%%%%%%%%%%%%%%%%%%%%%%%%%%%%%%%%%%%%%%%%%%%%%%%%%%%
%%%%%%%%%%%%%%%%%%%%%%%%%%%%%%%%%%%%%%%%%%%%%%%%%%%%%%%%%%%%%%%%%%%%%%%%%%%%%%%%%%%%%%%%%%%%%%%%%%%
\section{Introduction}
\label{sec:Introduction}
Nowadays, one of the most successful descriptions of Nature is based on Quantum Field Theories (QFT). Impressive calculations are being performed in order to extract highly-precise theoretical predictions, which must be confronted with the highly-accurate data collected from the current and future experiments \cite{Abada:2019lih,Abada:2019zxq,Benedikt:2018csr,Abada:2019ono,Blondel:2019vdq,Bambade:2019fyw,Roloff:2018dqu,CEPCStudyGroup:2018ghi}. Any tiny discrepancy in the theory-experiment comparison might open the path to huge discoveries regarding the fundamental building blocks of the Universe. Moreover, the challenge to achieve further improvements in the computational techniques behind QFT is originating several profound discoveries about the underlying mathematical structures of gauge theories. 

Due to the high-complexity of gauge theories and QFT in general, exact solutions are unknown for most of the experimentally relevant observables. In the context of high-energy physics, the perturbative approach turns out to be the most suitable, since it allows to express experimentally accessible quantities (such as cross-sections) in terms of power series: to increase the precision of the results, higher-orders must be included. These higher order contributions involve dealing with vacuum quantum fluctuations which are encoded through complicated multi-loop multi-leg Feynman diagrams.

In the last twenty years, there was an enormous progress towards more efficient loop calculations. Several techniques were developed, such as sector decomposition \cite{Binoth:2000ps,Smirnov:2008py,Carter:2010hi,Borowka:2017idc}, Mellin-Barnes transformations \cite{Blumlein:2000hw,Anastasiou:2005cb,Bierenbaum:2006mq,Gluza:2007rt,Freitas:2010nx,Dubovyk:2016ocz}, algebraic reduction of integrands \cite{Mastrolia:2011pr,Badger:2012dp,Zhang:2012ce,Mastrolia:2012an,Mastrolia:2012wf,Ita:2015tya,Mastrolia:2016dhn,Ossola:2006us}, integration-by-parts \cite{Chetyrkin:1981qh,Laporta:2001dd}, semi-numerical integration \cite{Francesco:2019yqt,Bonciani:2019jyb,Czakon:2008zk}, among other highly creative and powerful ideas \cite{Gnendiger:2017pys,Heinrich:2020ybq,TorresBobadilla:2020ekr}. In this direction, the Loop-Tree Duality (LTD) \cite{Catani:2008xa,Bierenbaum:2010cy,Bierenbaum:2012th,deJesusAguilera-Verdugo:2021mvg} constitutes a novel strategy to tackle higher-order calculations by opening loops into trees, thus recasting the virtual states into configurations that resemble real-radiation processes.

The purpose of LTD is twofold. On one side, expressing the virtual and real-radiation contributions on similar integration spaces allows to infer a natural way to combine them at integrand-level. This unified formalism produces an integrand-level representation of physical observables which is locally free of infrared singularities \cite{Hernandez-Pinto:2015ysa,Sborlini:2016fcj,Sborlini:2016gbr,Sborlini:2016hat}. 
On the other hand, there are tremendous simplifications in the description of the causal and singular structure of multi-loop Feynman integrals and scattering amplitudes, that leads to a more compact and numerically stable representation of the loop integrands in the Euclidean space of the loop three-momenta \cite{Buchta:2014dfa,Buchta:2015xda,Buchta:2015wna}. Several studies that took advantage of a simplified treatment of singularities within LTD were carried out \cite{Driencourt-Mangin:2017gop,Plenter:2019jyj,Plenter:2020lop,Driencourt-Mangin:2019sfl,Driencourt-Mangin:2019yhu}. Regarding the causal structure of scattering amplitudes, there are previous studies using different techniques \cite{Tomboulis:2017rvd,Runkel:2019yrs,Runkel:2019zbm}. Very recently, LTD was applied to remove unphysical threshold singularities and obtain a manifestly causal integrand-level definition of multi-loop scattering amplitudes \cite{Aguilera-Verdugo:2019kbz,Capatti:2019edf,Capatti:2019ypt,Verdugo:2020kzh,Ramirez-Uribe:2020hes,MANIFESTLYCAUSAL,Aguilera-Verdugo:2020kzc,Capatti:2020ytd}. All these LTD-based techniques were implemented in an automatized framework \cite{TorresBobadilla:2021dkq}.  

It is known that geometry and graph theory can be used to re-interpret the physical meaning of scattering amplitudes. Moreover, Cutkosky rules \cite{Cutkosky:1960sp,Bloch:2015efx} and Steinmann relations \cite{Cahill:1973qp,Cahill:1973px,Caron-Huot:2016owq} establish a deep connection among geometrical properties of Feynman diagrams (cuts or partitions) and the structure of discontinuities of the underlying amplitude. Inspired by these ideas, we investigated similar ideas with the purpose of reconstructing the whole amplitude at integrand-level using a manifestly causal representation. Even more, in a recent article, we explored the application of novel quantum algorithms to efficiently detect causal configurations in multi-loop diagrams, by identifying acyclic graphs with Grover's algorithm \cite{Ramirez-Uribe:2021ubp}.

The outline of this article is the following. In Sec. \ref{sec:LTDintro}, we briefly recall the basic ideas behind the Loop-Tree Duality theorem, and we introduce its connection with the causal structures in multi-loop multi-leg scattering amplitudes. Then, in Sec. \ref{sec:Geometry}, we establish the geometrical concepts required to describe Feynman integrals and multi-loop amplitudes. Introducing the concepts of multi-edges, vertices and the vertex matrix, we explain how to generate all the possible causal propagators involved in their causal representations in Sec. \ref{ssec:LambdasGenerados}. After that, in Sec. \ref{sec:Compatibility}, we describe a set of geometrical rules to unveil the causal structure of any amplitude. These rules explain how to combine compatible causal propagators, associated to different thresholds, leading to the concept of \emph{compatible causal entangled thresholds}. We present a detailed example based on four-vertex topologies in Sec. \ref{sec:FourPoint}. A discussion about more complicated configurations, including the causal structure of $N$-vertex topologies at one-loop, is given in Sec. \ref{sec:MoreExamples}. Finally, the conclusions and outlook are presented in Sec. \ref{sec:conclusions}.

%2021-07-03: DONE

%%%%%%%%%%%%%%%%%%%%%%%%%%%%%%%%%%%%%%%%%%%%%%%%%%%%%%%%%%%%%%%%%%%%%%%%%%%%%%%%%%%%%%%%%%%%%%%%%%%
%%%%%%%%%%%%%%%%%%%%%%%%%%%%%%%%%%%%%%%%%%%%%%%%%%%%%%%%%%%%%%%%%%%%%%%%%%%%%%%%%%%%%%%%%%%%%%%%%%%
\section{Loop-Tree Duality and Causality}
\label{sec:LTDintro}
In order to provide a proper description of multi-loop multi-leg scattering amplitudes, it is mandatory to identify and classify the kinematical variables involved. So, let's consider a generic $L$-loop $N$-point amplitude. In first place, we define $L$ primitive loop momenta $\{\ell_i\}_{i=1,\ldots,L}$ which correspond to the integration variables. 

Then, we group the momenta of the internal lines, $I$, associated to a Feynman diagram (or topology) into $n$ sets, according to their dependence on the primitive variables. In this way, the set $s$ contains all the internal momenta of the form
\beq
q_{i_s}=\sum_j \, \beta_j^{s} \ell_j \, + \, k_{i_s} \, ,
\label{eq:setDEF}
\eeq
where $k_{i_s}$ represents a linear combination of external momenta $\{p_r\}_{r=1,\ldots,N}$ and $\beta_j^{s} \in\{-1,0,1\}$. The linear combination of primitive momenta $\beta_j^{s} \ell_j$ remains fixed for each $i_s \in s$. Here, external momenta are considered outgoing and the short-hand notation $q_i \equiv i$ is used when there is only one element per set (i.e. $\#s = 1$ for every $s$).

%%%%%%%%%%%%%%%%%%%%%%%%%%%%%%%%%%%%%%%%%%%%%%%%%%%%%%%%%%%%%%%%%%%%%%%%%%%%%%%%%%%%%%%%
\begin{figure}[h]
    \centering
    \includegraphics[width=0.5\textwidth]{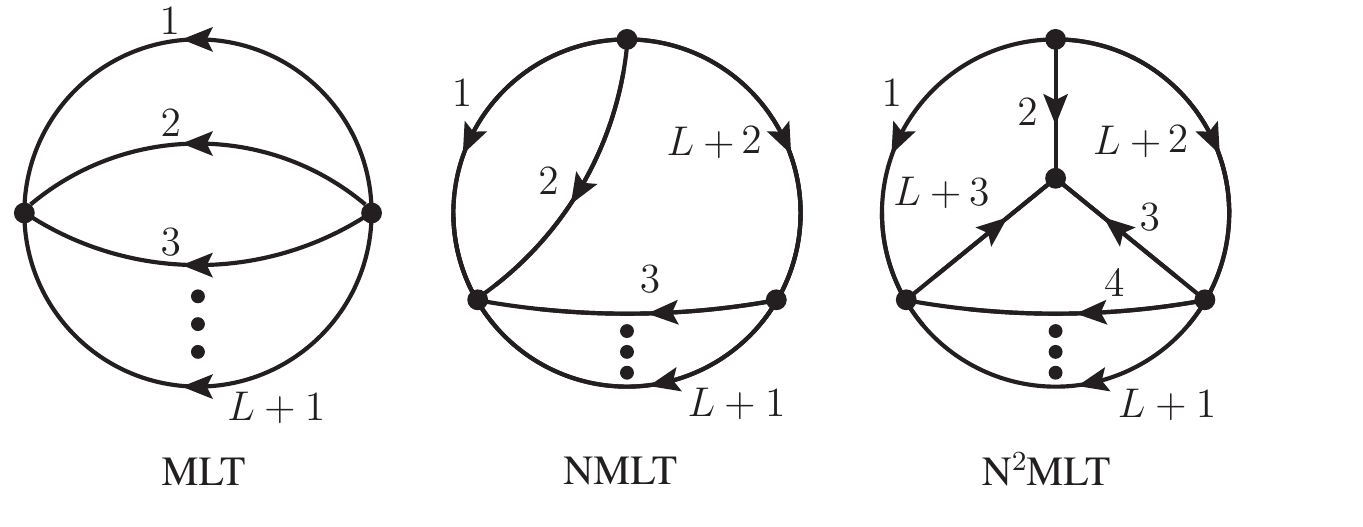}
    \caption{General examples of MLT (a), NMLT (b) and N$^2$MLT (c) topologies. We indicate with a single line each of the the sets defined by Eq. (\ref{eq:setDEF}). As explained in the text, we can attach an arbitrary number of external particles to each line, which corresponds to enlarging the sets $s$ by adding more propagators on them.}
    \label{fig:0}
\end{figure}
%%%%%%%%%%%%%%%%%%%%%%%%%%%%%%%%%%%%%%%%%%%%%%%%%%%%%%%%%%%%%%%%%%%%%%%%%%%%%%%%%%%%%%%%

At this point, we introduce the concept of \emph{Maximal Loop Topology} (MLT), which describes those diagrams or families of diagrams with the minimal number of sets for a given number of loops; i.e. $n-L =1$. This naturally defines a topological classification of diagrams through the so-called topological complexity, $\tilde{k} \equiv n - L$, as previously discussed in Refs. \cite{Verdugo:2020kzh,Aguilera-Verdugo:2020kzc,Ramirez-Uribe:2020hes,JesusAguilera-Verdugo:2020fsn}. In this way, MLT corresponds to $\tilde{k}=1$; the Next-to-Maximal Loop Topology (NMLT) to $\tilde{k}=2$, and so on. In Fig. \ref{fig:0}, we show generic examples of MLT (left), NMLT (center) and N$^2$MLT (right) diagrams. It is worth appreciating that the lines drawn in the diagrams represent sets, which might contain several external particles attached to them.

Once the notation has been established, we can use it to write any $L$-loop $N$-point diagram in the Feynman representation as 
\beq
\nonumber \mathcal{A}_{N}^{\left(L\right)}=\int_{\ell_{1},\hdots,\ell_{L}}\,  \mathcal{N}\left(\left\{ \ell_{i}\right\} _{L},\left\{ p_{j}\right\} _{N}\right)
\times G_{F}\left(1,\hdots,n\right)\,,
\label{eq:AmplitudGenerica}
\eeq
where $n=L+\tilde{k}$ is the total number of momentum sets, $\tilde{k}$ is the corresponding topological complexity of the diagram and ${\cal N}$ represents a generic numerator that depends on all the kinematical variables, i.e. any possible scalar product involving $\{p_j\}$ and $\{\ell_i\}$. In this expression, 
\beq
\int_{\ell} \, \equiv -\imath \, \mu^{4-d} \int\, \frac{d^{d}\ell}{(2\pi)^d} \, ,
\label{eq:measureNormal}
\eeq
corresponds to the standard $d$-dimensional loop integration measure. The product of Feynman propagators is given by
\begin{equation}
G_{F}\left(1,\ldots,n\right)=\prod_{i\in1\cup\cdots\cup n}\,(G_{F}(q_{i}))^{\alpha_{i}}\, ,
\label{eq:ProductoGFs}
\end{equation}
where $\alpha_i \in {\mathbb N}$. Individually, each Feynman propagator is expressed according to
\beq
G_F(q) = \frac{1}{(q_0-q_0^{(+)})(q_0+q_0^{(+)})} \,,
\label{eq:FeynmanPROP}
\eeq
where $q=(q_0,\vec{q})$ is the momentum carried by the particle and $q_0^{(+)}=\sqrt{(\vec q)^2+m^2-\imath 0}$ is the associated positive on-shell energy. This emphasizes that any internal line of a loop amplitude can be interpreted as the quantum superposition of two off-shell states flowing forward and backward in time, respectively. We will recall this interpretation later in Sec. \ref{sec:Compatibility}.

The LTD representation of Eq. (\ref{eq:AmplitudGenerica}) is obtained by integrating out one degree of freedom per loop through the Cauchy residue theorem. The application of the nested residues leads to a collection of diagrams with as many on-shell cuts as loops, in such a way that each loop diagram is open into a sum of non-disjoint trees. After adding together all the dual terms, the final result only involves same-sign combinations of on-shell energies in the denominator: these are the so-called \emph{causal propagators} \cite{Verdugo:2020kzh,Aguilera-Verdugo:2020kzc,Ramirez-Uribe:2020hes,JesusAguilera-Verdugo:2020fsn}. The causal propagators are associated to threshold discontinuities or singularities, as those predicted by the optical theorem and reconstructed through Cutkosky's rules \cite{Cutkosky:1960sp}.

As already investigated in previous articles, the LTD offers an excellent opportunity to disentangle the causal structure of scattering amplitudes. In particular, starting from Eq. (\ref{eq:AmplitudGenerica}) and computing the nested residues, we claim that \cite{Verdugo:2020kzh,Aguilera-Verdugo:2020kzc,Ramirez-Uribe:2020hes,Bobadilla:2021rmu}
\begin{align}
\nonumber & {\cal A}_{N}^{\left(L\right)}=\sum_{\sigma \in \Sigma}  \, \int_{\vec{\ell}_{1},\cdots,\vec{\ell}_{L}}\frac{{\cal N}_{\sigma}(\{q_{r,0}^{(+)}\},\{p_{j,0}\})}{x_{n}} \,
\\ & \times \prod_{i=1}^{k} \frac{1}{-\lambda_{\sigma(i)}}\, + \, (\sigma \leftrightarrow \bar{\sigma})  \, ,
\label{eq:MasterCausalFormula}
\end{align}
fully describes the causal structure of any multi-loop multi-leg Feynman diagram of order $k$. The order of a diagram is given by 
\beq
k = I - L\, ,
\label{eq:OrderDEF}
\eeq
namely, the number of remaining off-shell propagators after opening the loops into trees through $L$ iterated cuts. In Eq. (\ref{eq:MasterCausalFormula}), the causal propagators have the generic form
\beq
\lambda_j^\pm  \equiv \sum_{i \in o_j} q_{i,0}^{(+)} \pm k_j \, ,
\label{eq:LamDEF}
\eeq
where $k_j$ denotes a sum of external momenta and $o_j$ represents the internal lines that are on-shell in the associated threshold singularity. The set $\Sigma$ contains all the subsets of products of $k$ causal propagators, $\sigma$, which fulfill certain compatibility criteria. Also, $\bar{\sigma}$ is obtained from $\sigma$ through the replacement $\lambda_j^{\pm} \leftrightarrow \lambda_j^{\mp}$, namely by reversing simultaneously the momenta flow of all the internal lines. Additionally, we introduce the short-hand definitions
\beq
\int_{\vec \ell} \equiv \mu^{d-4} \int \frac{d^{d-1}\ell}{(2\pi)^{d-1}} \, , \ \ x_{n} =  \prod_{i\in 1\cup ...\cup n} 2 q_{i,0}^{(+)} \, ,
\eeq
which encode, respectively, the Euclidean dual integration measure and the normalization factor coming from the iterated application of Cauchy's residue theorem. Regarding the numerator, ${\cal N}_\sigma$ is given by the application of an operator depending on the subset $\sigma$ (whose explicit form might be obtained from a direct calculation of the residues) and only involves on-shell energies of the internal lines and the energies for external particles. It is worth appreciating that, for scalar integrals, we obtain ${\cal N}_\sigma \equiv 1$ for all $\sigma \in \Sigma$. Thus, a possible path to recover Eq. (\ref{eq:MasterCausalFormula}) consists in performing a reduction to scalar integrals and then computing their causal representations \cite{TorresBobadilla:2021dkq,Bobadilla:2021rmu}.

%2021-07-06: DONE

%%%%%%%%%%%%%%%%%%%%%%%%%%%%%%%%%%%%%%%%%%%%%%%%%%%%%%%%%%%%%%%%%%%%%%%%%%%%%%%%%%%%%%%%%%%%%%%%%%%
%%%%%%%%%%%%%%%%%%%%%%%%%%%%%%%%%%%%%%%%%%%%%%%%%%%%%%%%%%%%%%%%%%%%%%%%%%%%%%%%%%%%%%%%%%%%%%%%%%%
\section{Geometrical description of multi-loop amplitudes}
\label{sec:Geometry}
Multi-loop multi-leg scattering amplitudes are built from Feynman diagrams, i.e. geometrical structures described by graphs made of \emph{vertices} and \emph{lines}. The lines are understood as propagators that carry momenta and connect the different vertices. The vertices describe interactions among particles and impose momentum conservation involving internal and/or external particles. Propagators connect exactly two vertices, and there could be more than one propagator connecting two vertices. In that case, we substitute the sum of all the momenta flowing through propagators connecting to two vertices by a single \emph{multi-edge}; thus a \emph{multi-edge} corresponds to a bunch of lines, with the same origin and end, which are merged together. The set of all the multi-edges defines a basis $Q$, which is extended to include external momenta as well.

\begin{figure}[ht]
    \centering
    \includegraphics[width=0.46\textwidth]{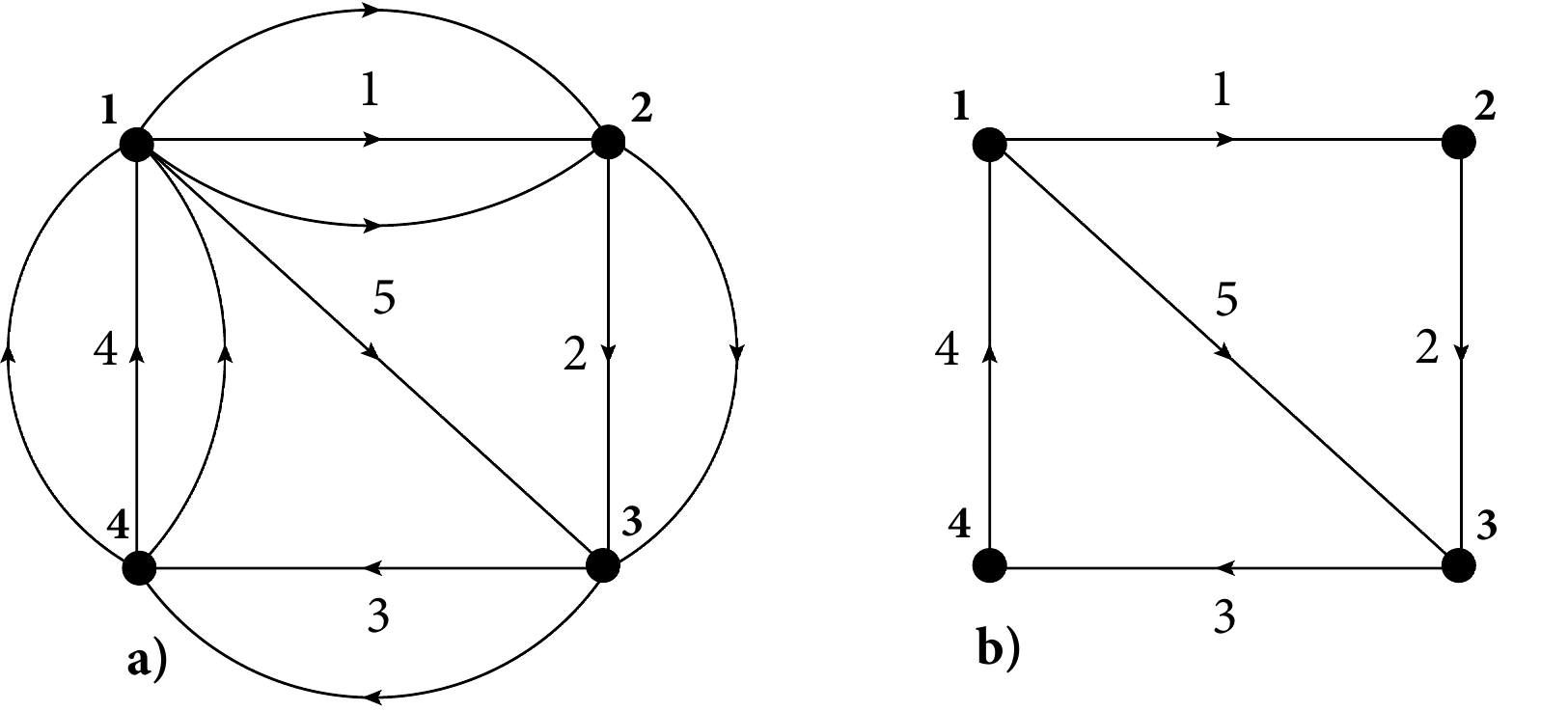}
    \caption{Comparison between the standard Feynman (left) and the reduced graph (right), for a four-vertex topology. Vertices are labelled with bold numbers. Whilst the diagram in \textbf{a} has 11 propagators and 8 loops, the reduced one in \textbf{b} has only 5 multi-edges and 2 graphical loops. Multi-edges 1 and 4 are the result of collapsing 3 lines; multi-edges 2 and 3 originate from 2 propagators; and multi-edge 5 is composed by a single line in the standard graph.}
    \label{fig:0a}
\end{figure}

Let's consider a multi-loop Feynman diagram with $N$ external particles, $L$ loops and $V$ interaction vertices. These vertices are connected through $I$ propagators, which can be reduced to $M$ multi-edges by merging those connecting the same vertices. Thus, the same Feynman diagram can be described in two equivalent ways:
\begin{itemize}
    \item Standard Feynman diagram: a graph with $L$ loops, $I$ propagators (lines) and $V$ vertices satisfying
    \beq
      V - 1 = I - L \, ,
      \label{eq:Euler1}
    \eeq
    i.e. Euler's formula.
    \item Reduced Feynman diagram: a graph with $V$ vertices connected by $M$ multi-edges, which satisfies an analogous conservation equation, 
    \beq
      V - 1 = M - \tilde{L} \, ,
      \label{eq:Euler2}
    \eeq
    with $\tilde{L}$ the number of \emph{graphical loops}.
\end{itemize}
Whilst in the standard representation each loop is associated to a loop integration, the \emph{graphical loops} only designate a topological characteristic of the reduced Feynman graph~\footnote{The graphical loops correspond to the concept of \emph{eloop} introduced in Ref. \cite{Ramirez-Uribe:2021ubp}.}. In order to clarify these concepts, we sketch the distinction between them in Fig. \ref{fig:0a} for a four-vertex topology. In the left side, we show the standard Feynman graph with $L=8$ and $I=11$. By merging lines into multi-edges, we obtain the reduced graph in the right side, which is composed by $M=5$ multi-edges and $\tilde{L}=2$ graphical loops. Additionally, if the multi-edge $e_1$ is the result of merging the lines $\{i_1,\ldots,i_r\}$, the associated energy is
\beq
q_{e_1,0} \equiv \sum_{j=1}^r q_{i_j,0}  \, ,
\label{eq:EnergyEdge}
\eeq
and the corresponding on-shell energy is given by 
\beq
q^{(+)}_{e_1,0} \equiv \sum_{j=1}^r q^{(+)}_{i_j,0}  \, .
\label{eq:OnshellEdge}
\eeq
The last definition is supported by the behaviour of the LTD representation of MLT-like insertions in multi-loop Feynman diagrams. As rigorously proven in Ref. \cite{JesusAguilera-Verdugo:2020fsn}, when several lines connect two vertices, they can be replaced by an equivalent propagator whose equivalent on-shell energy is the sum of the on-shell energies of each individual line \footnote{For more details, see Sec. 3 of Ref. \cite{JesusAguilera-Verdugo:2020fsn}.}. 

In general, we notice that combining Eq. (\ref{eq:OrderDEF}) with Eqs. (\ref{eq:Euler1})-(\ref{eq:Euler2}), we obtain
    \beq
      k = V - 1 \, ,
      \label{eq:EulerOrder}
    \eeq
which indicates that the order of a diagram is directly related to the number of vertices. It turns out that the reduced Feynman diagrams are more suitable to infer the causal structure of Feynman integrals (or amplitudes), as we will explain in the rest of the article. This assertion is also supported by similar studies based on algebraic properties of multi-loop Feynman diagrams constructed from vertices and multi-edges \cite{Bobadilla:2021rmu}.

%%%%%%%%%%%%%%%%%%%%%%%%%%%%%%%%%%%%%%%%%%%%%%%%%%%%%%%%%%%%%%%%%%%%%%%%%%%%%%%%%%%%%%%%
%%%%%%%%%%%%%%%%%%%%%%%%%%%%%%%%%%%%%%%%%%%%%%%%%%%%%%%%%%%%%%%%%%%%%%%%%%%%%%%%%%%%%%%%

Moving forward with the formulation of the geometrical framework, we define the basis of multi-edges and external momenta $Q$. For a given reduced Feynman diagram, we choose the ordering
\beq 
Q =\{Q_1,\ldots,Q_M;p_1,\ldots,p_{N-1}\} \, ,
\eeq 
where we implicitly impose $\sum p_i = 0$ due to momentum conservation. For each vertex, we have a unique momentum conservation equation: momentum is conventionally considered positive (negative) if it is outgoing (incoming). In this way, a vertex $v \in V$ is spanned over the basis $Q$, as a linear combination of multi-edge momenta with coefficients $\{\pm 1, 0\}$. For instance, given the vertex $v$ with multi-edges $Q_1$ and $Q_2$ outgoing, $Q_4$ incoming and the external outgoing momentum $p_2$ attached to it, we introduce the representation
\beqn
\nonumber v &=& (1^+,2^+,4^-;\hat{2}^+) 
\\ &\equiv& (1,1,0,-1,\ldots,0;0,1,\ldots,0) \, ,
\label{eq:Examplevertex}
\eeqn
with the short-hand notation $\hat{j}=p_j$. The vertex $v$ corresponds to the momentum conservation equation 
\beq  
v:  \longrightarrow Q_1+Q_2-Q_4+p_2=0 \, .
\eeq 
Then, we define the \emph{vertex matrix} ${\cal V}$ as the $V\times(M+N-1)$ rectangular matrix whose rows corresponds to the coordinates of all the vertices spanned on the basis $Q$. For example, if the first vertex is the one given in Eq. (\ref{eq:Examplevertex}), the generic structure of ${\cal V}$ would be
\begin{align}
{\cal V}=\left( \begin{array}{@{}c|c@{}}
   \begin{matrix}
      1 & 1 & 0 & -1 & \ldots \\
      0 & -1 & 1 & 0 & \ldots \\
        &  & \ldots & & \\
   \end{matrix} 
      & \begin{matrix}
      0 & 1 & 0 & \ldots \\
      0 & 0 & 1 & \ldots \\
       &  & \ldots & \\
   \end{matrix} 
\end{array} \right) \, ,
\label{eq:VertexMatrix}
\end{align}
where the vertical line separates internal (left) and external (right) momenta. 

All the kinematic information encoded in ${\cal V}$ is enough to unveil the causal structure of the underlying diagram as we will explain in the following. Global momentum conservation implies that 
\beq
{\rm Rank}({\cal V})=V-1 = k \, ,
\label{eq:RankV}
\eeq
since the information of the momenta entering (exiting) to (from) a given vertex is constrained by the whole system. We can make two direct observations. First, the rank of the vertex matrix agrees with the order of the diagram, by virtue of Eq. (\ref{eq:EulerOrder}). Second, we can implement a practical criterion to identify linear combinations of multi-edge momenta that are compatible with momentum conservation. Explicitly, given $q=\sum a_i Q_i$ we have a unique coordinate representation in the $Q$ basis and we can attach the corresponding row to the vertex matrix: if the extended matrix fulfils ${\rm Rank}({\cal V^*})\leq V-1$, then $q \equiv 0$ because of momentum conservation. This last property is useful to test the compatibility rules that define all the possible entangled thresholds.

%2021-07-06: DONE

%%%%%%%%%%%%%%%%%%%%%%%%%%%%%%%%%%%%%%%%%%%%%%%%%%%%%%%%%%%%%%%%%%%%%%%%%%%%%%%%%%%%%%%%%%%%%%%%%%%
%%%%%%%%%%%%%%%%%%%%%%%%%%%%%%%%%%%%%%%%%%%%%%%%%%%%%%%%%%%%%%%%%%%%%%%%%%%%%%%%%%%%%%%%%%%%%%%%%%%
\subsection{Generation of causal propagators}
\label{ssec:LambdasGenerados}
Here, we explain how the causal propagators are generated using concepts from graph theory. To do so, we consider reduced Feynman diagrams, i.e. we only rely on the information concerning vertices and multi-edges. Then, we define a \emph{binary partition} of the reduced graph as a non-trivial partition of the set of vertices $V=\{v_1,v_2,\ldots\}$ with two components: since one is the complement of the other, we identify each partition with the smaller subset of vertices. In general, we have
\beq
{\cal P}_V = \{ \{1\}, \{2\} , \ldots , \{1,2\}, \{1,3\}, \ldots\} \ ,
\label{eq:Partition}
\eeq
with the notation $j \equiv v_j$. ${\cal P}_V$ is the quotient set of all the possible subsets of $V$, constrained by the equivalence relation $r \equiv r^c$ where $r^c=V/r$.

Since Feynman diagrams fulfill momentum conservation (and reduced graphs inherit this property), a physical partition must also fulfil it. This means that vertices inside each element of the partition must be \emph{connected} through oriented multi-edges in a consistent way. We can encode this information by looking into the vertices contained in each element of ${\cal P}_V$ and its complement. By definition, a single vertex is self-connected. A set of vertices is connected if there exist multi-edges joining them two-by-two; since multi-edges converge into a vertex and they have a given orientation, this implies a consistent momentum flow in each vertex belonging to the partition. Thus, we define a \emph{connected} partition ${\cal P}_V^C$ as the subset of elements of $p \in {\cal P}_V$ such that $p$ and $p^c$ are connected. The connection is defined by the existence of at least one path that allows to go from one vertex to any other inside the partition.

Let us use the reduced graph in Fig. \ref{fig:0a} to illustrate the concept of connection. The set of all the possible binary partitions is given by
\beq
{\cal P}_V = \{ \{1\}, \{2\} , \{3\}, \{4\} , \{1,2\}, \{1,3\},  \{1,4\} \} \ ,
\label{eq:PartitionFIG0}
\eeq
but not all of them are connected. This is the case of $p=\{1,3\}$, which includes the vertices 1 and 3 connected through the multi-edge 5. However, its complement, $p^c=\{2,4\}$, involves disconnected vertices: there is not a multi-edge joining 2 and 4. In consequence, $p=\{1,3\} \not\in {\cal P}_V^C$. Also, with this definition, all the remaining binary partitions in Eq. (\ref{eq:PartitionFIG0}) are connected.

The connected binary partitions originated from a reduced graph are important since they codify the threshold structure of the corresponding Feynman amplitude. Then, we need to establish the relation among partitions and causal denominators. In order to do this, given $p \in {\cal P}_V^C$, we define its \emph{conjugated causal propagator} as the sum of all the energies of the associated multi-edge momenta connecting the vertices inside the partition. If $\beta_j \in \{\pm 1,0\}$ and $\gamma_j \in \{1,0\}$, a generic conjugated causal propagator is given by
\beq
\bar{\lambda}_p = \sum_j \beta_j \, Q_{j,0} + \sum_{i=1}^{N-1} \gamma_i \, p_{i,0}\, ,
\label{eq:ConjugatedCausal}
\eeq
where the coefficients $\beta_j$ reflects the freedom to choose the propagator momenta flow in the Feynman representation and $\bar{\lambda}_p=0$ when momentum conservation is fulfilled in all the vertices associated to the partition $p$. External particles are always labelled as outgoing, even if they carry negative energy (which is equivalent to say that they are actually incoming particles). It is worth appreciating that this definition is consistent, i.e. we recover the same expression for $\bar{\lambda}_{p^c}$ (i.e. by considering $p^c$), because of global momentum conservation. Also, we notice that $\bar{\lambda}_p$ has a strong physical meaning: it is the total momenta flowing from (or to) a given \emph{binary partition} of a reduced Feynman diagram.

Once the \emph{conjugated causal propagators} are defined, we can generate all the possible \emph{causal propagators}. They originate from the overlap of momentum conservation and the nested application of Cauchy's residue theorem, leading to causal same-sign combinations of on-shell energies. Explicitly, we introduce the transformation
\beq
\bar{\lambda}_p \to \pm \lambda^{\pm}_p = \pm \sum_j |\beta_j| \, Q^{(+)}_{j,0} + \sum_{i=1}^{N-1} \gamma_i \,  p_{i,0} \, ,
\label{eq:CausalTransformation}
\eeq  
which is equivalent to fix a partition, evaluate the nested residues (i.e. replace loop energy components by positive on-shell energies, $Q_{i,0}^{(+)}$) and consistently align all the momenta involved. 

To conclude this Section, we shall recall the discussion given in Eq. (\ref{eq:FeynmanPROP}): propagators involve the superposition of two off-shell modes, which implies that they can be aligned in two possible directions once they become on-shell. Then, we appreciate that $\lambda^{\pm}_p$ is determined modulo a global sign. Besides that, we use the convention in Eq. (\ref{eq:CausalTransformation}) since it implies $\lambda^+=\lambda^-$ in the absence of external momenta attached to some vertices, thus simplifying the expressions. Finally, we will equivalently denote the causal propagators by the vertices involved in the associated connected binary partition, i.e. $\lambda_p \equiv \{v_{i_1},\ldots, v_{i_r}\}$.

%2021-07-06: DONE

%%%%%%%%%%%%%%%%%%%%%%%%%%%%%%%%%%%%%%%%%%%%%%%%%%%%%%%%%%%%%%%%%%%%%%%%%%%%%%%%%%%%%%%%%%%%%%%%%%%
%%%%%%%%%%%%%%%%%%%%%%%%%%%%%%%%%%%%%%%%%%%%%%%%%%%%%%%%%%%%%%%%%%%%%%%%%%%%%%%%%%%%%%%%%%%%%%%%%%%
\section{Causality and compatibility conditions}
\label{sec:Compatibility}
As mentioned before, the connected binary partitions of a reduced Feynman diagram and all the associated causal propagators are in strict correspondence. Each causal propagator corresponds to a different physical threshold, as discussed in Refs. \cite{Aguilera-Verdugo:2019kbz,Verdugo:2020kzh}. On the other hand, multi-loop multi-leg Feynman diagrams involve a superposition of several thresholds, that might occur when different combinations of internal states become on-shell. Thus, from a geometrical point of view, the threshold or causal structure of a given Feynman diagram is determined by a set of specific combinations of connected binary partitions. Each possible term in a representation like Eq. (\ref{eq:MasterCausalFormula}) corresponds to \emph{entangled causal thresholds}, using the concepts defined in Ref. \cite{Aguilera-Verdugo:2020kzc}. So, in this Section, we explain how to determine all the allowed entangled causal thresholds imposing geometrical selection rules to combine the different causal propagators.

In first place, we notice that:
\begin{enumerate}
\setcounter{enumi}{0}
    \item \emph{Each possible entangled combination of causal denominators involves the on-shell energies of all the propagators}.
\end{enumerate}
This condition is related to the fact that we are cutting the diagram into trees, and classifying these trees according to their threshold structure. At one-loop, this agrees with the original Cutkosky's rules \cite{Cutkosky:1960sp}. Beyond one-loop, there are more possibilities to decompose the topologies into trees, but the idea remains the same: identify all the \emph{causal compatible} tree-level structures inside a diagram. Here, two tree-level blocks are \emph{causal compatible} if they fulfill momenta conservation and if they can be combined in such a way that the momenta exiting one block consistently enter into the other. 

With the previous ideas, we can re-interpret the generalization of Cutkosky's rules in terms of vertices and causal propagators. We claim that the fundamental objects to build the causal representation are the elements of ${\cal P}_V^C$, since they contain information about the momenta flow and the propagators that can be simultaneously set on-shell in each contribution. Each $\lambda_p$ is associated to a connected binary partition of the original reduced Feynman diagram in which internal lines are all entering or exiting the partition. Thus, we motivate the following two compatibility criteria:  
\begin{enumerate}
\setcounter{enumi}{1}
    \item \emph{Absence of crossing}: Two compatible causal propagators $\lambda_p$ and $\lambda_q$ must fulfill that the associated connected sets of vertices are disjoint or one is totally included in the other. Explicitly, if $p$ is smaller than $q$, then $p \subset q$ or $p \subset q^c$.
    \item \emph{Consistent multi-edge momenta orientation}: A set of causal propagators is compatible if the associated $\bar{\lambda}_p$ can be consistently oriented, i.e. if all the multi-edges contribute to the entangled cuts with the same orientation.
\end{enumerate}
Regarding the last criterion, let's remember that a causal propagator $\lambda^{\pm}_p$ corresponds to a binary partition which splits the reduced diagram into two connected pieces. At this point, the causal denominators can be thought as aligned contributions obtained from $\bar{\lambda}_p$ by selecting either negative or positive energy modes. Then, this criterion implies that all the multi-edges crossing the cut must have the same direction, i.e. they are all entering or exiting the partition. In order to achieve this, we might have to reverse some multi-edges. If a diagram involves several cuts or causal thresholds, we have to look for a consistent orientation of all the multi-edges of the reduced diagram. In consequence, it could happen that when combining several $\lambda_p$, some multi-edges can not be consistently aligned (for instance, because $q_j \equiv -q_j$ is required). 

It turns out that imposing criterion 3 leads to a reduced Feynman diagram without cycles, i.e. an acyclic oriented graph. We call \emph{causal orientation} of the multi-edges to a configuration of momenta flow which converts the reduced graph into an acyclic oriented graph. Thus, given a reduced graph, we can first find all the causal orientations and then look for those combinations of $\lambda_p$ which fulfil criteria 1-3. This problem can be tackled by studying the eigenvalues of the adjacency matrix, which is build from the vertex matrix introduced in Eq. (\ref{eq:VertexMatrix}) \footnote{Recent developments on quantum algorithms to speed-up the identification of causal orientations were presented in Ref. \cite{Ramirez-Uribe:2021ubp}.}.

\begin{figure}[h]
    \centering
    \includegraphics[width=0.45\textwidth]{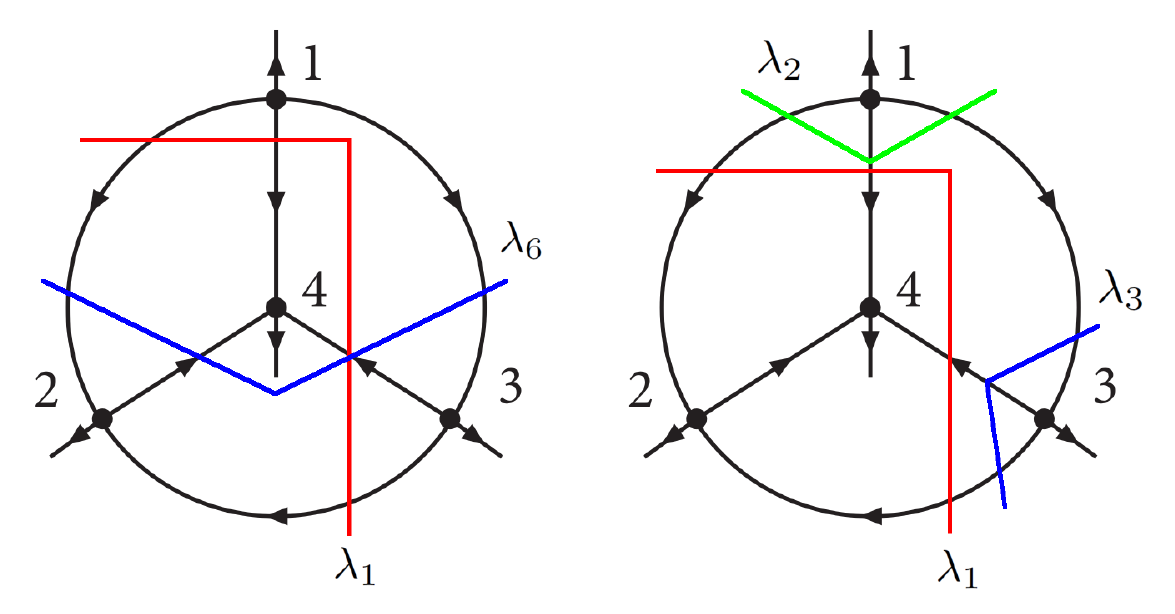}
    \caption{(Left) Example of two incompatible causal propagators with non-disjoint sets of vertices. (Right) Example of a forbidden combination of causal propagators due to incompatible momenta orientation.}
    \label{fig:1}
\end{figure}

To illustrate the application of criteria 1-3, let's consider a typical 3-loop Mercedes-Benz diagram as shown in Fig. \ref{fig:1}. We assume that there is only one line per multi-edge, in such a way that the reduced and the standard Feynman diagram are the same. First, we define 
\beq
\lambda_1 \equiv \{1,3\} \ , \ \lambda_2 \equiv \{1\} \ , \ \lambda_3 \equiv \{3\}  \ , \ \lambda_6 \equiv \{1,4\} \, ,
\label{eq:LambdasMercedes}
\eeq
which corresponds to the identification among partitions of vertices and conjugated causal propagators. On the left graph, $\{1,3\} \not\subset \{1,4\}$ and $\{1,3\} \not\subset \{2,3\}=\{1,4\}^c$ since these are non-disjoint sets of vertices: criterion 2 is not fulfilled here. On the right graph, the entangled threshold corresponding to $(\lambda_1 \, \lambda_2 \, \lambda_3)$ is considered. If we fix the orientation of the lines exiting the partition $\{1,3\}\equiv \lambda_1$, the multi-edge connecting the vertices 1 and 3 can not be consistently oriented. Thus, this combination of causal thresholds is forbidden from the causal representation by virtue of criterion 3.

After applying criterion 3 and ordering the multi-edges to represent acyclic directed graphs, we need to keep the information about those multi-edges that were reversed, i.e. the causal orientations. This is important to apply the transformation given in Eq. (\ref{eq:CausalTransformation}) and properly determine whether it corresponds to select $\lambda_p^+$ or $\lambda_p^-$. In fact, we have the following rule:
\begin{enumerate}
\setcounter{enumi}{3}
\item \emph{Causal propagator orientation}: Given a connected binary partition in a causally oriented reduced graph, if the external momenta and the oriented multi-edges are both outgoing, then $\bar{\lambda}_p \to \lambda_p^+$. Otherwise, $\bar{\lambda}_p \to \lambda_p^-$.
\end{enumerate}
Summarizing the procedure described up to now and applying the criteria 1-4 to any reduced Feynman graph, we can obtain the set of all the allowed entangled causal thresholds describing that diagram. We will denote this set as $\bar{\Sigma}$. 

However, some of the entangled thresholds defined by criteria 1-4 might be degenerated due to global momentum conservation. This is true when the number of multi-edges is not enough to fully constrain the flow of the different cuts involved in an entangled threshold. Equivalently, the degeneration takes place if the number of multi-edges is not maximal for the associated topology. This observation motivates the following definition: a reduced Feynman diagram is a \emph{maximally connected graph} (MCG) if all the vertices are connected to each other. In other words, the associated adjacency matrix can be transformed into an upper-triangular $\#V \times \# V$ matrix with all the entries equal to 1. After exploring several topologies, we find the last rule:
\begin{enumerate}
\setcounter{enumi}{4}
\item \emph{Removing the threshold degeneration}: Given a non-maximally connected graph, we select a pair of disconnected vertices, $i$ and $j$, and force the condition $p_i=-p_j\equiv q_{M+1}$. We repeat the procedure for all the disconnected vertices, till we generate a maximally connected graph. Then, we force the validity of criterion 3 including the fictitious multi-edges $\{q_{M+1},\ldots,q_{M+R-1}\}$.
\end{enumerate}
This criterion determines the set $\Sigma \subset \bar{\Sigma}$ in Eq. (\ref{eq:MasterCausalFormula}) and fixes a causal representation of the diagram. Of course, this also shows that there might be several equivalent causal representations for a given Feynman graph, being all of them related due to momentum conservation. Only when the reduced Feynman graph is maximally connected, we have $\Sigma = \bar{\Sigma}$. This condition is also true when the diagram is \emph{next-to-maximally connected}, in the sense that only two vertices are disconnected. For more general cases, criterion 5 might take a more complicated explicit form; we defer for future studies these configurations and the exploration of their corresponding geometrical properties.

%%%%%%%%%%%%%%%%%%%%%%%%%%%%%%%%%%%%%%%%%%%%%%%%%%%%%%%%%%%%%%%%%%%%%%%%%%%%%%%%%%%%%%%%%%%%%%%%%%%
%%%%%%%%%%%%%%%%%%%%%%%%%%%%%%%%%%%%%%%%%%%%%%%%%%%%%%%%%%%%%%%%%%%%%%%%%%%%%%%%%%%%%%%%%%%%%%%%%%%
\section{Application to four-vertex topologies}
\label{sec:FourPoint}
After presenting the abstract geometrical formalism in the previous Sections, we will provide a concrete application. But first, we want to recall that all the information about the causal structure of any Feynman diagram is encoded within the vertex matrix. Independently of the number of loops and external legs, the computational complexity of a Feynman diagram is given by the number of vertices and how they are connected. Rephrasing a bit the last two sentences, all the selection criteria described in Sec. \ref{sec:Compatibility} can be implemented through operations performed on ${\cal V}$.

\begin{figure}[h]
    \centering
    \includegraphics[width=0.46\textwidth]{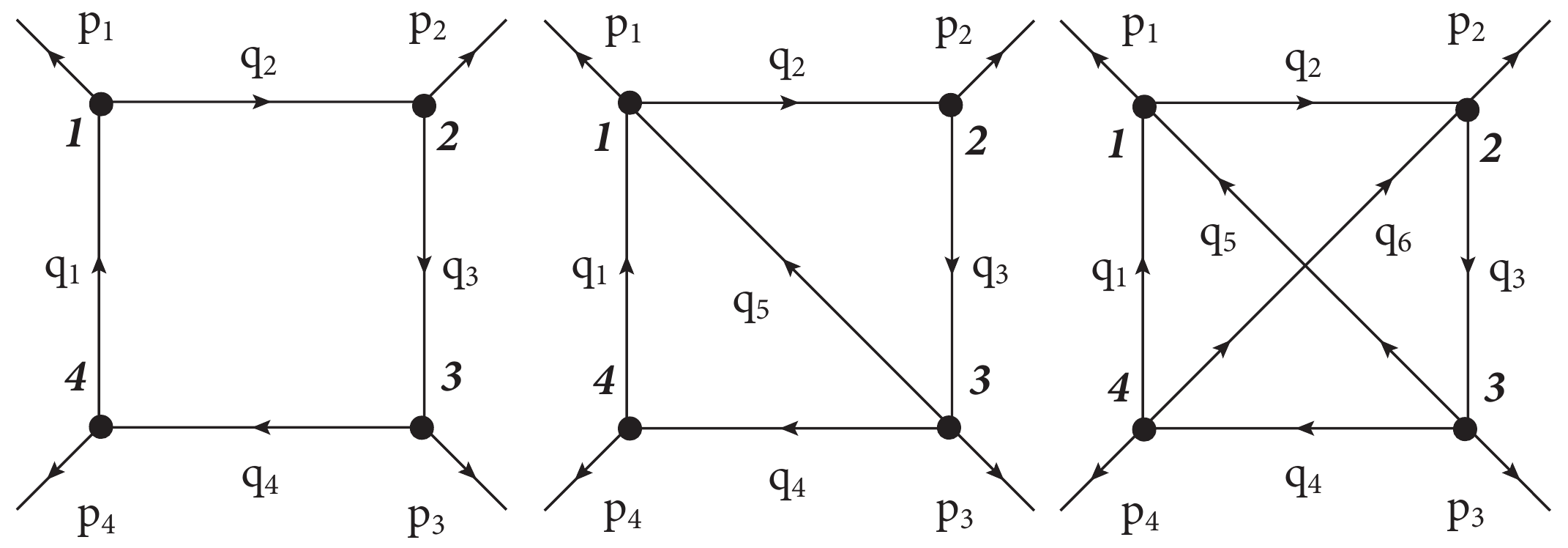}
    \caption{Four-vertex topologies with 4, 5 and 6 multi-edges, respectively. These topologies involve four external particles.}
    \label{fig:3}
\end{figure}

So, let us illustrate our framework for the case of the scalar four-vertex topologies. We start with the case with 4 multi-edges depicted in Fig. \ref{fig:3} (left), i.e. a one-loop four-point function. For the sake of simplicity, we consider that each multi-edge is composed by a single propagator (i.e. the reduced and standard Feynman graphs are equal), whose associated momenta are
\beqn
\nonumber && q_1 = \ell_1 \, , \  q_2 = \ell_1 - p_1 \, , \ q_3 = \ell_1 - p_1 - p_2 \, ,
\\ && q_4 = \ell_1 - p_1 - p_2 - p_3 \, , 
\eeqn
and $\{p_i\}_{i=1,\ldots,4}$ are the external momenta fulfilling $\Sigma p_i=0$. The basic set of momenta is
\beq
Q=\{1,2,3,4;\hat{1},\hat{2},\hat{3}\} \, ,
\eeq
and $\hat{4}=-\hat{1}-\hat{2}-\hat{3}$ because of global momentum conservation. The vertices are
\begin{align}
\nonumber &v_1 = (1^-,2^+;\hat{1}^+) \, , \quad v_2 = (2^-,3^+;\hat{2}^+) \, ,
\\ &v_3 = (3^-,4^+;\hat{3}^+) \, , \quad v_4 = (4^-,1^+;\hat{4}^+) \, ,
\label{eq:verticesBOX}
\end{align}
which leads to the vertex matrix
\begin{align}
{\cal V}=\left( \begin{array}{@{}c|c@{}}
   \begin{matrix}
      -1 & 1 & 0 & 0 \\
      0 & -1 & 1 & 0 \\
      0 & 0 & -1 & 1 \\
      1 & 0 & 0 & -1
   \end{matrix} 
      & \begin{matrix}
      1 & 0 & 0 \\
      0 & 1 & 0 \\
      0 & 0 & 1 \\
      -1 & -1 & -1 
   \end{matrix} 
\end{array} \right) \, .
\label{eq:vertexMatrixBOX}
\end{align}
By looking into the connected partitions of vertices, we find
\beq
{\cal P}_V^C = \{\{1\},\{2\},\{3\},\{4\},\{1,2\},\{1,4\}\} \, ,
\label{eq:PartitionBOX}
\eeq
where we only keep the equivalence classes determined by $p \equiv p^c$. We notice that $\{1,3\}\equiv \{2,4\} \not\in {\cal P}_V^C$ because both of them involve non-adjacent vertices which can not be connected. Associated to each element of ${\cal P}_V^C$, we have the corresponding conjugated causal propagator, i.e.
\begin{align}
\nonumber & \bar{\lambda}_1 = (-1,1,0,0;1,0,0) \equiv \{1\} \, ,
\\ \nonumber & \bar{\lambda}_2= (0,-1,1,0;0,1,0) \equiv \{2\} \, ,
\\ \nonumber & \bar{\lambda}_3 = (0,0,-1,1;0,0,1) \equiv \{3\} \, ,
\\ \nonumber & \bar{\lambda}_4 = (1,0,0,-1;-1,-1,-1) \equiv \{4\}\, ,
\\ \nonumber &\bar{\lambda}_5 = (-1,0,1,0;1,1,0) \equiv \{1,2\} \, , 
\\ & \bar{\lambda}_6 = (0,1,0,-1;0,-1,-1) \equiv \{1,4\} \, ,
\label{eq:ConjugatedCausalBOX1L}
\end{align}
being the internal (external) coordinates located to the left (right) of the semicolon. Applying the transformation described in Eq. (\ref{eq:CausalTransformation}) to Eq. (\ref{eq:ConjugatedCausalBOX1L}), we obtain all the causal propagators for this topology. Explicitly, we have
\beqn
\nonumber \lambda_1^{\pm}&=&q_{1,0}^{(+)}+q_{2,0}^{(+)}\pm p_{1,0} \, ,
\\ \nonumber \lambda_2^{\pm}&=&q_{2,0}^{(+)}+q_{3,0}^{(+)}\pm p_{2,0} \, ,
\\ \nonumber \lambda_3^{\pm}&=&q_{3,0}^{(+)}+q_{4,0}^{(+)}\pm p_{3,0} \, ,
\\ \nonumber \lambda_4^{\pm}&=&q_{1,0}^{(+)}+q_{4,0}^{(+)}\pm (p_{1,0}+p_{2,0}+p_{3,0}) \, ,
\\ \nonumber \lambda_5^{\pm}&=&q_{1,0}^{(+)}+q_{3,0}^{(+)}\pm (p_{1,0}+p_{2,0}) \, ,
\\ \lambda_6^{\pm}&=&q_{2,0}^{(+)}+q_{4,0}^{(+)}\pm (p_{2,0}+p_{3,0}) \, ,
\label{eq:PropagatorsCausalBOX1L}
\eeqn
by replacing the energy component of each multi-edge with the associated aligned positive on-shell energies, i.e. $q_{i,0}^{(+)}$.

Once we generated all the connected binary partitions, we need to identify the allowed entangled thresholds. First, we notice that the order of this diagram is $k=3$, because it is a four vertex topology and we apply Eq. (\ref{eq:EulerOrder}). Since there are 6 causal propagators, the number of potential combinations of thresholds is 20. The application of criteria 1 and 2 reduces the possibilities to only 16. More combinations are discarded after considering criteria 3-4, which involve a compatible ordering of the multi-edges. Thus, given an entangled threshold, we test all the possible orderings of multi-edges that leads to an acyclic graph; we retain only those configurations where multi-edges are aligned when entering/exiting all the binary partitions. This information allows to define an \emph{ordering matrix} that is used to distinguish between $\lambda_i^+$ and $\lambda_i^-$. Thus, applying the transformation rules in Eq. (\ref{eq:CausalTransformation}), we obtain
\beqn
\nonumber && \bar{\Sigma} = \{(1^+,2^-,3^+),(1^+,2^-,4^+),(1^+,2^-,6^-),
\\ \nonumber && (1^+,3^+,4^+),(1^+,3^-,5^+),(1^+,3^-,6^-),(1^+,4^+,5^+),
\\ \nonumber && (2^+,3^-,4^-),(2^+,3^-,5^+),(2^+,4^+,5^+),(2^+,4^+,6^+),
\\ && (3^+,4^+,6^+)\} \, ,
\label{eq:BOXpermutations}
\eeqn
which is the set of all the compatible entangled thresholds. Here, we use the short-hand notation $i^\pm \equiv \lambda_i^{\pm}$. Notice that $\#(\bar{\Sigma})=12$.

At this point, we appreciate that setting $\Sigma = \bar\Sigma$ and ${\cal N}_\sigma \equiv 1$ for all $\sigma \in \Sigma$ in Eq. (\ref{eq:MasterCausalFormula}) does not agree with the result of the explicit nested residue calculation. This is because there are degenerated entangled thresholds: in fact,
\beq
\{(1,2,3),(1,3,4)\} \equiv \{(1,2,4),(2,3,4)\} \ ,
\label{eq:Degenerados}
\eeq
because of momentum conservation. By including these configurations we over-count the effect of some entangled thresholds \footnote{An alternative approach consists in calculating symmetry factors for all the possible entangled thresholds. In this example, if we set ${\cal N}_\sigma = 1/2!$ for the configurations in Eq. (\ref{eq:Degenerados}) and ${\cal N}_\sigma = 1$ for the others, we obtain the proper result. We defer for a future publication the exploration of this alternative representation and its properties.}. Thus, we need to break the degeneration by applying criterion 5. So, we can choose to close the loop by joining vertices 1 and 3, or 2 and 4. In the first case, we obtain
\beqn
\nonumber && \Sigma_1 = \{(1^+,2^-,4^+),(1^+,2^-,6^-),(1^+,3^-,5^+),
\\ \nonumber && (1^+,3^-,6^-),(1^+,4^+,5^+),(2^+,3^-,4^-),(2^+,3^-,5^+),
\\ && (2^+,4^+,5^+),(2^+,4^+,6^+),(3^+,4^+,6^+)\} \, ,
\label{eq:BOXpermutations1}
\eeqn
whilst in the second
\beqn
\nonumber && \Sigma_2 = \{(1^+,2^-,3^+),(1^+,2^-,6^-),(1^+,3^+,4^+),
\\ \nonumber && (1^+,3^-,5^+),(1^+,3^-,6^-),(1^+,4^+,5^+),(2^+,3^-,5^+),
\\ && (2^+,4^+,5^+),(2^+,4^+,6^+),(3^+,4^+,6^+)\} \, ,
\label{eq:BOXpermutations2}
\eeqn
where both sets contain 10 elements. Following the functional form presented in Eq. (\ref{eq:MasterCausalFormula}), causal representations of the scalar one-loop four-vertex topology are given by
\begin{align}
&&{\cal A}_{\text{4-vertex}}^{\left(\text{1-loop}\right)}=\int_{\vec{\ell}_{1}}\frac{1}{x_{4}} \,\sum_{\sigma \in \Sigma_r} \, \prod_{i=1}^{3} \frac{-1}{\lambda_{\sigma(i)}}\, \ + (\lambda_i^+ \leftrightarrow \lambda_i^-) \,,
\label{eq:BOXformula}
\end{align}
with 
\beq
x_4^{-1}=16 q_{1,0}^{(+)} q_{2,0}^{(+)} q_{3,0}^{(+)} q_{4,0}^{(+)} \, ,
\eeq
using either $r=1$ or $r=2$, from Eqs. (\ref{eq:BOXpermutations1}) and (\ref{eq:BOXpermutations2}) respectively.

%20210711: DONE

%%%%%%%%%%%%%%%%%%%%%%%%%%%%%%%%%%%%%%%%%%%%%%%%%%%%%%%%%%%%%%%%%%%%%%%%%%%%%%%%%%%%%%%%%%%%%%%%%%%
%%%%%%%%%%%%%%%%%%%%%%%%%%%%%%%%%%%%%%%%%%%%%%%%%%%%%%%%%%%%%%%%%%%%%%%%%%%%%%%%%%%%%%%%%%%%%%%%%%%
\subsection{Maximally and next-to-maximally connected four-vertex topologies}
\label{ssec:MAX4VERTICES}
Then, let's consider the remaining four-vertex topologies with 5 and 6 multi-edges. Starting from the two-loop four-vertex topology in Fig. \ref{fig:3} (center), we generate all the possible connected binary partitions. We immediately realize that they are the same as for the one-loop four-vertex case; i.e. the set ${\cal P}_V^C$ is also given by Eq. (\ref{eq:PartitionBOX}). However, the functional form of the corresponding causal propagators changes because of the additional multi-edge. Explicitly, we have
\beqn
\nonumber \lambda_1^{\pm}&=&q_{1,0}^{(+)}+q_{2,0}^{(+)}+q_{5,0}^{(+)}\pm p_{1,0} \, ,
\\ \nonumber \lambda_2^{\pm}&=&q_{2,0}^{(+)}+q_{3,0}^{(+)}\pm p_{2,0} \, ,
\\ \nonumber \lambda_3^{\pm}&=&q_{3,0}^{(+)}+q_{4,0}^{(+)}+q_{5,0}^{(+)}\pm p_{3,0} \, ,
\\ \nonumber \lambda_4^{\pm}&=&q_{1,0}^{(+)}+q_{4,0}^{(+)}\pm (p_{1,0}+p_{2,0}+p_{3,0}) \, ,
\\ \nonumber \lambda_5^{\pm}&=&q_{1,0}^{(+)}+q_{3,0}^{(+)}+q_{5,0}^{(+)}\pm (p_{1,0}+p_{2,0}) \, ,
\\ \lambda_6^{\pm}&=&q_{2,0}^{(+)}+q_{4,0}^{(+)}+q_{5,0}^{(+)}\pm (p_{2,0}+p_{3,0}) \, ,
\label{eq:PropagatorsCausalBOX2L}
\eeqn
Again, there are 20 potential candidates to be entangled thresholds (6 causal propagators to be grouped in sets of $k=3$ elements). Applying criteria 1-2, we discard 8 configurations. When imposing criteria 3-4, we realize that the presence of an additional multi-edge (w.r.t. the one-loop case) leads to stricter constraints and 2 other configurations are eliminated. Thus, we obtain
\beqn
\nonumber && \bar{\Sigma} = \{(1^+,2^-,4^+),(1^+,2^-,6^-),(1^+,3^-,5^+),
\\ \nonumber && (1^+,3^-,6^-),(1^+,4^+,5^+),(2^+,3^-,4^-),(2^+,3^-,5^+),
\\ && (2^+,4^+,5^+),(2^+,4^+,6^+),(3^+,4^+,6^+)\} \, ,
\label{eq:BOXpermutations2L}
\eeqn
which is the same set $\Sigma_1$ presented in Eq. (\ref{eq:BOXpermutations1}). This is not a coincidence: this two-loop topology is a next-to-maximally connected graph (NMCG), and removing the multi-edge $q_5$ leads to the one-loop box described in the previous discussion. Thus, using Eq. (\ref{eq:MasterCausalFormula}), we get
\begin{align}
&&{\cal A}_{\text{4-vertex}}^{\left(\text{2-loop}\right)}=\int_{\vec{\ell}_{1}\vec{\ell}_{2}}\frac{1}{x_{5}} \,\sum_{\sigma \in \Sigma} \, \prod_{i=1}^{3} \frac{-1}{\lambda_{\sigma(i)}}\, \ + (\lambda_i^+ \leftrightarrow \lambda_i^-) \,,
\label{eq:BOXformula2L}
\end{align}
as a causal representation for the scalar two-loop four-vertex topology, where
\beq
x_5^{-1}=32 q_{1,0}^{(+)} q_{2,0}^{(+)} q_{3,0}^{(+)} q_{4,0}^{(+)} q_{5,0}^{(+)} \, ,
\eeq
and $\Sigma \equiv \bar\Sigma$ given by Eq. (\ref{eq:BOXpermutations2L}).

Finally, let's consider the four-vertex topology with 6 multi-edges, shown in Fig. \ref{fig:3} (right). This topology is straightforwardly a maximally connected graph; all the vertices are connected. Using the identification between conjugated causal propagators and the elements of the connected binary partitions ${\cal P}_V^C$, we have
\begin{align}
\nonumber & \bar{\lambda}_1 \equiv \{1\} \, , \  \bar{\lambda}_2 \equiv \{2\} \, , \  \bar{\lambda}_3 \equiv \{3\} \, , \  \bar{\lambda}_4 \equiv \{4\} \, , 
\\ & \bar{\lambda}_5 \equiv \{1,2\} \, , \  \bar{\lambda}_6 \equiv \{1,4\} \, , \ \bar{\lambda}_7 \equiv \{1,3\} \, .
\end{align}
Notice that $\{1,3\}\equiv \{2,4\}$ was not a connected binary partition for the two previous topologies, but it contributes to this one. The introduction of additional edges allows to define a path connecting the vertices 1 and 3, as well as 2 and 4. Explicitly, the causal propagators are given by
\beqn
\nonumber \lambda_1^{\pm}&=&q_{1,0}^{(+)}+q_{2,0}^{(+)}+q_{5,0}^{(+)}\pm p_{1,0} \, ,
\\ \nonumber \lambda_2^{\pm}&=&q_{2,0}^{(+)}+q_{3,0}^{(+)}+q_{6,0}^{(+)}\pm p_{2,0} \, ,
\\ \nonumber \lambda_3^{\pm}&=&q_{3,0}^{(+)}+q_{4,0}^{(+)}+q_{5,0}^{(+)}\pm p_{3,0} \, ,
\\ \nonumber \lambda_4^{\pm}&=&q_{1,0}^{(+)}+q_{4,0}^{(+)}+q_{6,0}^{(+)}\pm (p_{1,0}+p_{2,0}+p_{3,0}) \, ,
\\ \nonumber \lambda_5^{\pm}&=&q_{1,0}^{(+)}+q_{3,0}^{(+)}+q_{5,0}^{(+)}+q_{6,0}^{(+)}\pm (p_{1,0}+p_{2,0}) \, ,
\\ \nonumber \lambda_6^{\pm}&=&q_{2,0}^{(+)}+q_{4,0}^{(+)}+q_{5,0}^{(+)}+q_{6,0}^{(+)}\pm (p_{2,0}+p_{3,0}) \, ,
\\ \lambda_7^{\pm}&=&q_{1,0}^{(+)}+q_{2,0}^{(+)}+q_{3,0}^{(+)}+q_{4,0}^{(+)}\pm (p_{1,0}+p_{3,0}) \, 
\label{eq:PropagatorsCausalBOX3L}
\eeqn
Regarding possible causal representations for this topology, we have 35 entangled thresholds (i.e. all the possible subsets of 3 causal propagators taken from the 7 available ones). Imposing criteria 1-2 eliminates 19 combinations, and we further reduce this set by requiring criteria 3-4 to be fulfilled. The remaining 12 allowed causal entangled thresholds are given by 
\beqn
\nonumber \bar\Sigma &&= \{(1^+,2^-,6^-),(1^+,2^-,7^+),(1^+,3^-,5^+),
\\ \nonumber && (1^+,3^-,6^-),(1^+,4^+,5^+),(1^+,4^+,7^+),
\\ \nonumber && (2^+,3^-,5^+),(2^+,3^-,7^-),(2^+,4^+,5^+),
\\ && (2^+,4^+,6^+),(3^+,4^+,6^+),(3^+,4^+,7^+)\} \, .
\label{eq:BOXpermutations3L}
\eeqn
Again, we notice that \emph{the} causal structure of this scalar four-vertex diagram is 
\begin{align}
&&{\cal A}_{\text{4-vertex}}^{\left(\text{3-loop}\right)}=\int_{\vec{\ell}_{1}\vec{\ell}_{2}\vec{\ell}_{3}}\frac{1}{x_{6}} \,\sum_{\sigma \in \Sigma} \, \prod_{i=1}^{3} \frac{-1}{\lambda_{\sigma(i)}}\, \ + (\lambda_i^+ \leftrightarrow \lambda_i^-) \,,
\label{eq:BOXformula3L}
\end{align}
 with
\beq
x_6^{-1}=64 q_{1,0}^{(+)} q_{2,0}^{(+)} q_{3,0}^{(+)} q_{4,0}^{(+)} q_{5,0}^{(+)} q_{6,0}^{(+)} \, ,
\eeq
and $\Sigma \equiv \bar\Sigma$ given by Eq. (\ref{eq:BOXpermutations3L}). As in the case of the two-loop box, the set of causal entangled thresholds is not degenerated (i.e. criterion 5 is immediately fulfilled). For this reason, we emphasize that it is the only possible causal representation compatible with Eq. (\ref{eq:MasterCausalFormula}).

To conclude this Section, we highlight that the procedure followed here is only based on geometrical concepts. In the three examples reported, we compared the reconstructed causal representations with the integrand-level result of the LTD representation (which was obtained through the explicit computation of the nested residues). A perfect agreement was found in all the cases.

%20210711: DONE

%%%%%%%%%%%%%%%%%%%%%%%%%%%%%%%%%%%%%%%%%%%%%%%%%%%%%%%%%%%%%%%%%%%%%%%%%%%%%%%%%%%%%%%%%%%%%%%%%%%
%%%%%%%%%%%%%%%%%%%%%%%%%%%%%%%%%%%%%%%%%%%%%%%%%%%%%%%%%%%%%%%%%%%%%%%%%%%%%%%%%%%%%%%%%%%%%%%%%%%
\section{Causal structure of multi-vertex diagrams}
\label{sec:MoreExamples}
Taking a step forward in complexity, we consider the causal structure of scalar $N$-vertex reduced diagrams. The possible number of multi-edges, $M$, fulfills
\beq
N \leq M \leq \frac{(N-1)N}{2} \, ,
\eeq
where the upper bound corresponds to the maximally connected topologies. By direct computation, we explored several examples and computed the possible causal entangled thresholds.

In first place, we studied the generation of all the possible causal propagators $\lambda_p^{\pm}$ starting from the sets of binary connected partitions for different topologies. In particular, we compared the results obtained with the geometric algorithm with the expressions reported in Refs. \cite{Aguilera-Verdugo:2020kzc,Ramirez-Uribe:2020hes}, finding complete agreement.

Then, we center into the generation of the allowed causal entangled thresholds for different topologies. In particular, we studied the causal structure of scalar maximally and next-to-maximally connected graphs. Based on explicit computational exploration, we conjecture that their causal structures are given by
\begin{align}
&&{\cal A}_{\text{N-vertex}}^{\text{(N)MCG}}=\int_{\vec{\ell}_{1}\ldots \vec{\ell}_{L}}\frac{1}{x_{M}} \,\sum_{\sigma \in \Sigma} \, \prod_{i=1}^{N-1} \frac{-1}{\lambda_{\sigma(i)}}\, \ + (\lambda_i^+ \leftrightarrow \lambda_i^-) \,,
\label{eq:NpointScalarMaximal}
\end{align}
where $\Sigma \equiv \bar\Sigma$, i.e. the set of allowed causal entangled thresholds after applying criteria 1-4. It is worth appreciating that the number of loops, $L$, only enters in this formula through the integration measure and the explicit dependence of each $q_{i,0}^{(+)}$ (with $i\in \{1,\ldots,M\}$). Thus, this supports the initial claim that the causal structure is independent of the number of loops and propagators (only depends of the vertices and multi-edges, as also reported in Ref. \cite{Bobadilla:2021rmu}).

Finally, we explored the opposite limit, i.e. the possible causal representations of $N$-vertex diagrams at one-loop. It might sound counter-intuitive that one-loop topologies are more complicated to describe than multi-loop ones. The point is that the geometrical reconstruction algorithm exploits the restrictions imposed by the momentum flow among vertices. Thus, maximally connected graphs are very constrained and the criteria 1-4 lead to a set of causal entangled thresholds that is not degenerated. One-loop $N$-point amplitudes are described by \emph{minimally connected graphs}, and there is an over-counting of configurations. To remove the degeneration, we need to apply criterion 5 and force global momentum conservation. However, we found an alternative way to select a non-degenerated set of causal thresholds. Explicitly, for one-loop diagrams, we realized that a particular choice of the selections rules dictated by criterion 5 is:
\begin{enumerate}
\setcounter{enumi}{4}
\item \emph{Removing the threshold degeneration (one-loop case)}: Given an entangled threshold, we keep the ordering matrix obtained after criteria 3-4 and apply it to the corresponding conjugated causal propagators $\bar\lambda_p$. Then, we adjust the direction of the multi-edges in such a way that they are all outgoing from the associated partition. If they were entering the partition, we reverse both of their fluxes ($q_i \to -q_i$) and the direction of the external momenta attached to that partition. If the external momenta are consistently aligned, we include the configuration in $\Sigma$; otherwise, we exclude it. 
\end{enumerate}
Criterion 5 can be implemented by constructing a matrix whose rows are the coordinates of $\bar\lambda_p$ in the basis $Q$, and applying transformations on the rows and columns. Also, we notice that this modified version of criterion 5 implicitly uses global momentum conservation, since the direction of the external momenta is fixed by the condition $p_N = - \sum p_i$.

In this way, causal representations for scalar $N$-point one-loop functions can be also described by Eq. (\ref{eq:NpointScalarMaximal}) with $M=N$, $L=1$ and the set $\Sigma$ determined with criteria 1-5. We checked the validity of our claim with several scalar one-loop $N$-point functions ($N \leq 9$), finding a complete agreement with the results computed through nested residues. 

%20210711: DONE

%%%%%%%%%%%%%%%%%%%%%%%%%%%%%%%%%%%%%%%%%%%%%%%%%%%%%%%%%%%%%%%%%%%%%%%%%%%%%%%%%%%%%%%%%%%%%%%%%%%
%%%%%%%%%%%%%%%%%%%%%%%%%%%%%%%%%%%%%%%%%%%%%%%%%%%%%%%%%%%%%%%%%%%%%%%%%%%%%%%%%%%%%%%%%%%%%%%%%%%
\section{Conclusions}
\label{sec:conclusions}
In this article, we presented a geometrical study of the entangled causal structure of multi-loop multi-leg Feynman integrals and amplitudes. We showed that all the information concerning the causal decomposition of a topology is encoded in the \emph{vertex matrix}. Diagrams with different numbers of legs and loops, but sharing the same number of vertices and multi-edges, exhibit a similar causal structure. We exemplified this situation studying the four-vertex topologies with 4, 5 and 6 multi-edges, respectively. Moreover, we introduced a classification of the different topologies based on graph theory: the order of the diagram is given by $k=V-1$ ($V$ number of vertices) and it indicates the number of causal thresholds that must be entangled.

In order to unveil the causal structure, we implemented an algorithm to generate all the possible causal propagators associated to a given (reduced) Feynman diagram by inquiring into a more fundamental object: the connected binary partitions of vertices. Then, we developed 4 criteria to select all the allowed causal entangled thresholds. We relied only on combinatorics, geometry and graph theory. Specifically, we imposed restrictions that allow to split the diagram into disconnected and non-overlapping tree-level graphs with a consistent momentum flow through their multi-edges. To consistently align the momentum flow of the multi-edges, we implemented transformations in the vertex matrix to identify all the possible acyclic directed graphs associated to the reduced Feynman diagram.  

For the case of maximally and next-to-maximally connected graphs (i.e. those where all the vertices are connected or only one multi-edge is missing), it turns out that criteria 1-4 lead to a set of entangled thresholds that seems to reconstruct the causal structure conjectured in Eq. (\ref{eq:MasterCausalFormula}). However, for topologies with $M < (N-1)N/2$, some causal entangled thresholds are degenerated due to global momentum conservation. In fact, criteria 1-4 only use information regarding the vertices and how they are \emph{internally} connected through multi-edges. The momentum conservation associated to external particles can be understood as \emph{adding an additional vertex in which all external momenta converge}.

Thus, we introduced an additional selection criterion and explored its consequences. In particular, we studied the case of $N$-vertex one-loop diagrams (which we called \emph{minimally connected graphs}) and obtained a simplified recipe to eliminate the degeneration due to momentum conservation. Our results were in complete agreement with the ones obtained through the explicit calculation of nested residues. 

The findings regarding our geometrical approach suggest an strong connection with the algebraic framework proposed in Ref. \cite{Bobadilla:2021rmu}. In that work, causal representations were obtained from maximally connected graphs and applying algebraic reduction relations. Thus, it would be highly interesting to understand the interplay between these two frameworks, since they tackle the same problem with two very different approaches.

Finally, the discoveries reported in this article establish an interesting connection between geometry, algebra and causality in Quantum Field Theories. Moreover, the effects of imposing global momentum conservation on scattering amplitudes deserve to be better understood, since they lead to several equivalent causal representations. Inquiring more on these findings might open new and more powerful paths to explore and compute higher-orders, breaking the precision frontier and unveiling the hidden mathematical structures in QFT.

%%%%%%%%%%%%%%%%%%%%%%%%%%%%%%%%%%%%%%%%%%%%%%%%%%%%%%%%%%%%%%%%%%%%%%%%%%%%%%%%%%%%%%%%%%%%%%%%%%%%%%%%%%%%%%%%%%%%%%%%%%
\section*{Acknowledgments}
We gratefully acknowledge the inspiring discussions with G. Rodrigo, R. Hern\'andez Pinto and W. J. Torres Bobadilla. We thank the author of Ref. \cite{Bobadilla:2021rmu} for making his results available for comparison, prior to the publication. We also thank N. S. Ram\'irez-Uribe and J. J. Aguilera-Verdugo for additional cross-checks. This article is based upon work from COST Action PARTICLEFACE CA16201, supported by COST (European Cooperation in Science and Technology). \href{https://www.cost.eu/}{www.cost.eu}.%This work is supported by the COST Action CA16201 PARTICLEFACE.

%\bibliography{refs}


\end{document}